\documentclass[preprint,12pt]{elsarticle}

\usepackage{amssymb}

 \usepackage{lineno}
\usepackage{subfig}
\usepackage{xcolor}

\journal{Astroparticle Physics}

\begin{document}
\begin{frontmatter}



\title{Study of Dispersion of Mass Distribution of Ultra-High Energy Cosmic Rays using a Surface Array of \\Muon and Electromagnetic Detectors}


\author[label 1]{Jakub V\'icha}
\cortext[mycorrespondingauthor]{Corresponding author}
\ead{vicha@fzu.cz}
\author[label 1]{Petr Tr\'avn\'i\v{c}ek}
\author[label 2]{Dalibor Nosek}
\author[label 1]{Jan Ebr}
\address[label 1]{Institute of Physics of the Academy of Sciences of the Czech Republic, \\Na Slovance 2, 182 21 Prague 8}
\address[label 2]{Faculty of Mathematics and Physics, Charles University in Prague, \\V Hole\v{s}ovi\v{c}k\'{a}ch 2, 180 00 Prague 8}

\begin{abstract}
We consider a hypothetical observatory of ultra--high energy cosmic rays consisting of two surface detector arrays that measure independently electromagnetic and muon signals induced by air showers. Using the constant intensity cut method, sets of events ordered according to each of both signal sizes are compared giving the number of matched events. Based on its dependence on the zenith angle, a parameter sensitive to the dispersion of the distribution of the logarithmic mass of cosmic rays is introduced. The results obtained using two post--LHC models of hadronic interactions are very similar and indicate a weak dependence on details of these interactions.
\end{abstract}
 

\end{frontmatter}


\section{Introduction}
\label{sec:introduction}
At any cosmic ray observatory of ultra--high energy cosmic rays (UHECR\footnote{Cosmic rays with energy above $10^{18}$~eV.}) such as the largest ground arrays operating currently, the Telescope Array \cite{TA:enespec} and the Pierre Auger Observatory \cite{AUGER:enespec}, the energy reconstruction requires a correction of the measured ground signal to account for the zenith angle of an incoming primary particle initiating the air shower. This correction reflects different amounts of air masses penetrated by air showers that reach the detector with different zenith angles. It can be obtained using the data--driven constant intensity cut (CIC) method \cite{Hersil:cicmet} or using a Monte Carlo (MC) based estimation. These two approaches result in different energy calibrations in the case of a mixed mass composition \cite{ICRC}. The MC--based approach produces biased results for reconstructed energies with respect to the zenith angle when averaged over masses of primary particles. On the other hand, the CIC approach does not suffer from these shortcomings and provides correct reconstructed energies on average.

In this study we adopted the CIC method to explore its capabilities to gain additional information about the primary mass composition. The CIC method is based on measured data. It assumes that the flux of incoming particles is isotropic above a given energy. This implies a flat distribution of $\cos^2\Theta$ where $\Theta$ denotes the zenith angle of recorded showers. The CIC approach selects a set of $N_{\rm Cut}$ events with the highest signal in each bin of $\cos^2\Theta$. The number $N_{\rm Cut}$ corresponds to an UHECR flux above a certain energy. The relationship between the minimal signal of the selected events and the mean value of $\cos^2\Theta$ in a bin defines the attenuation curve. When fluorescence detectors are available to calibrate the ground array, only the relative shape of the attenuation curve, known as the CIC curve, is important in the energy reconstruction procedure. In this study the CIC curve is normalized such that it is equal to one for the zenith angle $\Theta=38^{\circ}$. It is worth noting that the construction of CIC curves is stable. Even if a very strong source is present at the highest energies, the shape of the determined CIC curves is only a little distorted \cite{ICRC}.     

In general, the ground arrays used for the detection of UHECR showers are sensitive to secondary muons, to the electromagnetic component (EM) of an air shower or to their combinations. There are several previous works, e.g. \cite{FluctAndCIC,KascadeCIC}, studying the primary mass composition and its influence on the CIC method and vice versa. Usually, the detected muon and EM signals are utilized to separate primary mass groups, see e.g.~\cite{KascadeSpectrum}, or to determine the average mass number of a set of air showers, see e.g.~\cite{KascadeLnA}. The dispersion of the distribution of the primary mass is more difficult to obtain. The precise fluorescence measurement of the distributions of the depth of shower maximum is used, see e.g \cite{AUGER:LongXmaxPaper,WorkingGroup}, albeit with a low duty cycle. The combination of measurements of the mean value and the dispersion of the distribution of the primary mass is discussed in \cite{KampertUnger,AUGER:Interpretation}. However, these analyses suffer from a strong dependence on models of hadronic interactions. Recently, a new method estimating the spread of masses in the UHECR primary beams has been presented \cite{CorrelationPaper}. Unlike our analysis, this method is based on the simultaneous measurements of the depth of shower maximum and the muon shower size.

In this study we consider a hypothetical observatory comprising two independent arrays of particle detectors (full duty cycles) with different responses to shower muons and to the EM component. The CIC approach applied simultaneously to both types of signals is used to calculate the number of events with the highest energies matched in both detectors. The zenith angle behaviour of this number provides us with information regarding the spread of primary masses. 

The main purpose of the article is to present a method how to obtain information about the spread of primary masses from the data collected simultaneously by different types of surface detectors. We use average features of CORSIKA \cite{CORSIKAref} showers simulated at an energy of $10^{19}$~eV as inputs to a fast and simplified simulation of signals in both detectors caused by showers over a wide range of primary energies. The application to the measured data would require a precise knowledge of the detection process. In our analysis the detailed detector responses are not included. Instead, detector imperfections are represented by a simple Gaussian smearing of signals.

The article is organized as follows. In Section~2, we deal with a simulation of the detection of air showers with two arrays of detectors sensitive to muon and EM components. Reference signals inferred from CORSIKA showers at an energy of $10^{19}$~eV are described in Section~2.1. Simplified simulations of shower signals over a wide range of energies are presented in Section~2.2. In Section~2.3, a parameter sensitive to the dispersion of the distribution of primary mass is introduced. Our results are presented and discussed in Section~3, and summarized in the section following that.
 
\section{Simulation of UHECR Detection}
\label{sec:SimDescription} 
To address the details of the CIC method a large simulated data sample is necessary, ideally $\sim 10^{6}$ simulated showers, that is comparable with the achievable statistics of the largest UHECR experiment ever built. To avoid excessive computational requirements, we generated a set of showers induced by proton (p), helium (He), nitrogen (N) and iron (Fe) primaries with an energy of $10^{19}$~eV. These showers were produced by CORSIKA ver.~7.37 (Section~2.1). From signals that they produce in both arrays we derived their fluctuations and correlations. Finally, we constructed attenuation curves for both types of signals. These curves were utilized in the simplified simulation of the muon and EM signals induced by showers over a wide range of energies (Section~2.2).

A hypothetical observatory with independent muon and EM detectors was placed at ground level, 1400~m~a.s.l. (880~g/cm$^{2}$ of vertical depth). The signal of the muon detector was assumed to be proportional to the ground density of muons with a threshold energy $E_{\rm Th}=500$~MeV. The signal of the EM detector was modeled to be proportional to the ground density of EM particles with $E_{\rm Th}$~=~1~MeV. These detector responses were motivated by responses of thin scintillators shielded by 250~g/cm$^{2}$ of mass overburden (muon detector) and thin unshielded scintillators (EM detector).

\subsection{Reference Shower Signals}
\label{RefShowerSignals}
In our study the reference CORSIKA showers were simulated at a fixed energy of $10^{19}$~eV. Although the signal fluctuations and the shapes of the attenuation curve depend slightly on the shower energy, our final results are not affected by such variations. To describe low energy interactions, the FLUKA model \cite{FLUKA} was used. The high energy interactions were simulated using the two most up--to--date models tuned to the LHC data: QGSJet~II--04 \cite{QGSJETII04} and EPOS--LHC \cite{epos,epos2}. About 60 showers were produced for each primary, each model of high energy interactions and for each of seven zenith angles between 0$^{\circ}$ and 45$^{\circ}$ maintaining equal steps in $\cos^{2}(\Theta)$. 

\begin{figure}[h!]
  \centering
\includegraphics[width=1.0\textwidth]{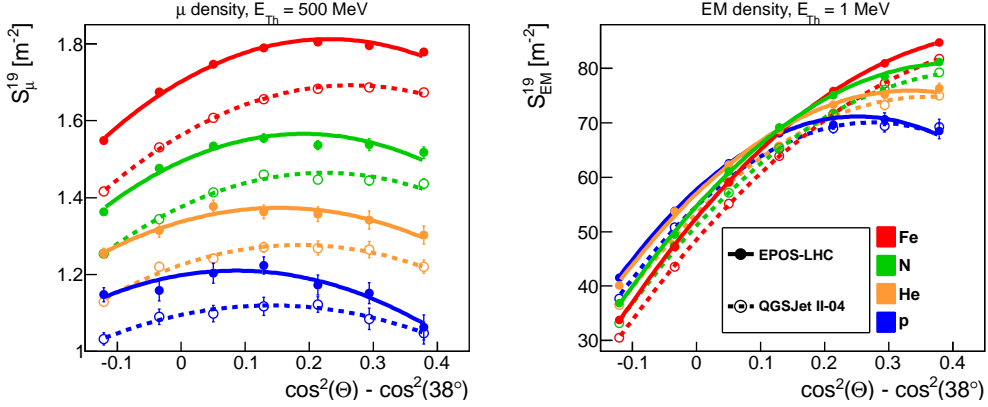}
\caption{Attenuation curves. Reference signals of CORSIKA showers of energy $10^{19}$~eV are fitted with quadratic functions of $\cos^{2}(\Theta)$ for the muon detector (left panel) and the EM detector (right panel) in the range $\Theta \in$~${\langle}0^{\circ},45^{\circ}{\rangle}$. Two models of hadronic interactions and four primary species are distinguished by types of lines and colors, respectively.}
  \label{CICfit}
\end{figure}

The reference muon and EM signals, $S_{\mu}^{19}$ and $S_{\rm EM}^{19}$, were determined as the densities of corresponding particles averaged over these 60 showers at a distance of 1000~m from the shower core. 
Both these reference signals were fitted by quadratic functions of $\cos^{2}(\Theta)$ (attenuation curves) with precisions at the level of a few percent.
The muon and EM attenuation curves are depicted in Fig.~\ref{CICfit}. 
They depend on the type of the primary particles. 
The EM signal obeys a stronger dependence on the zenith angle than the muon signal. 

\begin{figure}[h!]
  \centering
\includegraphics[width=1.0\textwidth]{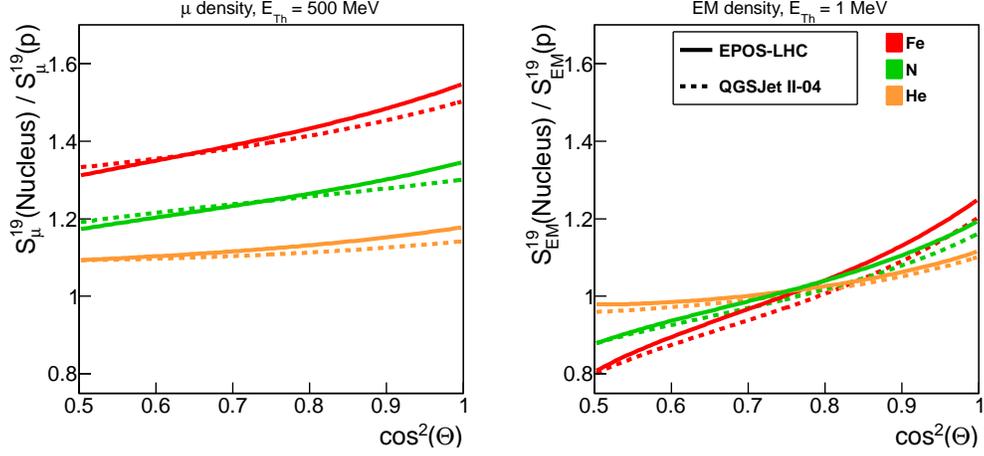}
  \caption{Ratios of shower signals. Nucleus to proton ratios for the muon (left panel) and EM (right panel) signals are plotted as a function of cos$^{2}(\Theta)$. CORSIKA simulations at energy $10^{19}$~eV are used. Two models of hadronic interactions and three primary nuclei are distinguished by types of lines and colors, respectively.}
  \label{NucleiToProtonRatios}
 \end{figure}

In Fig.~\ref{NucleiToProtonRatios}, the ratios of ground signals induced by primary He, N and Fe nuclei to the proton induced signal are depicted for the muon (left panel) and EM detector (right panel). 
Whereas the ratio for $S_{\mu}^{19}$ is greater than one for all zenith angles and increases with the mass number of the primary particle, the ratio for $S_{\rm EM}^{19}$ decreases more steeply with zenith angle than in the case of $S_{\mu}^{19}$, and is even smaller than one beyond $\Theta\sim 30^{\circ}$. 
This different behavior of $S_{\mu}^{19}$ and $S_{\rm EM}^{19}$ with the zenith angle for different primary particles plays the main role in the considerations described in Section \ref{sec:MatchedFraction}.

\begin{figure}[h!]
  \centering
	\subfloat{\includegraphics[width=0.5\textwidth]{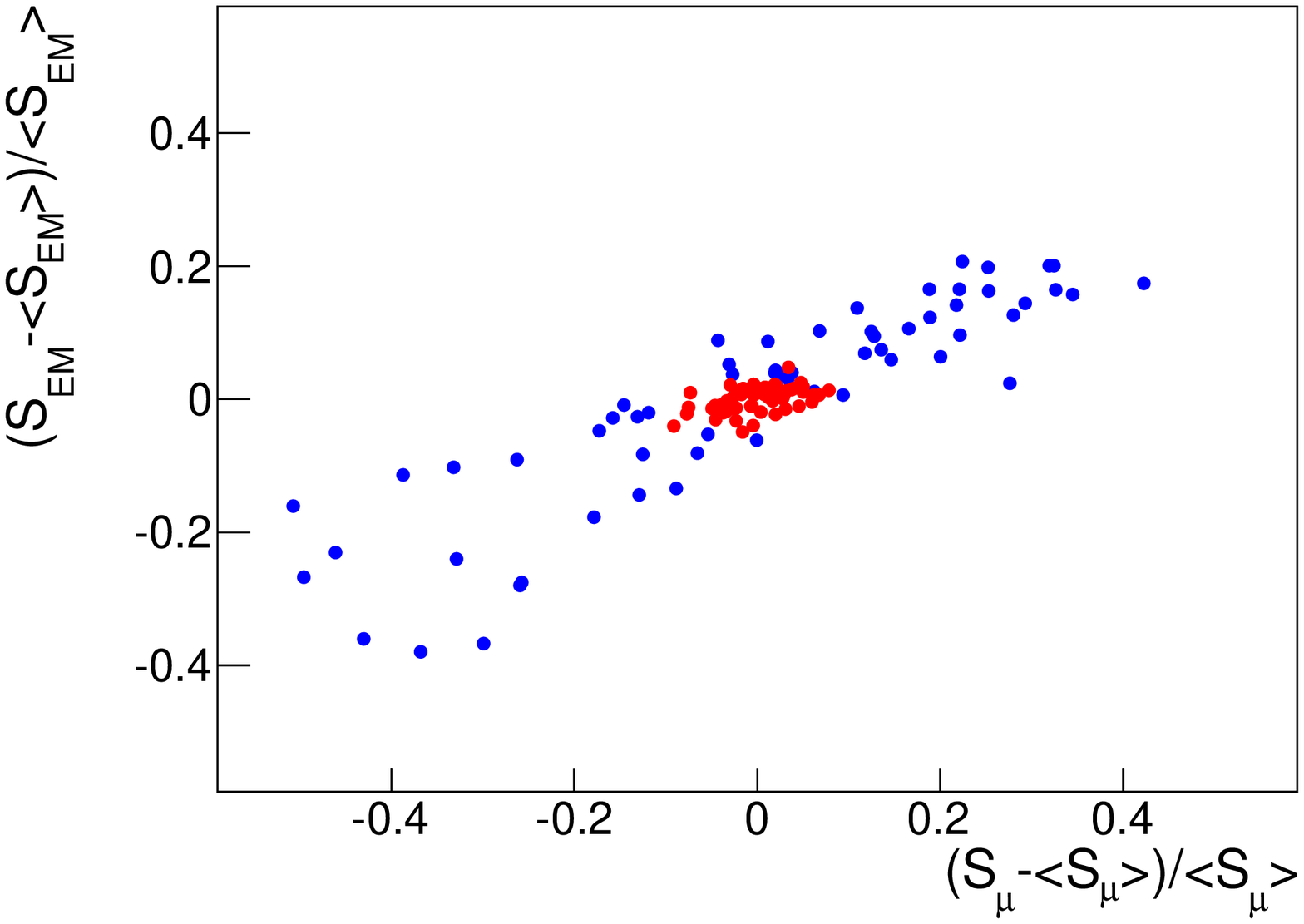}}
	\subfloat{\includegraphics[width=0.5\textwidth]{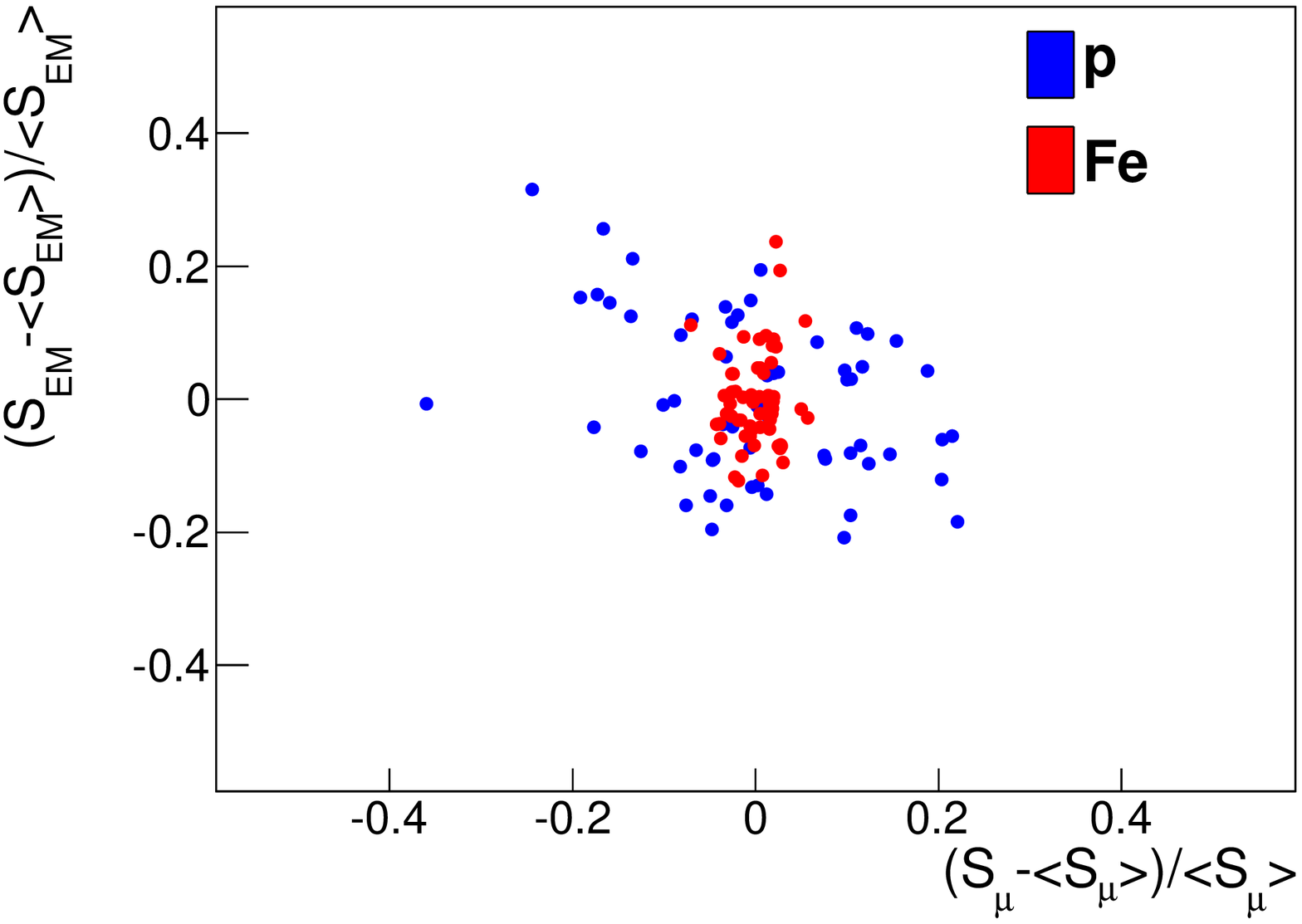}}
  \caption{Spread of the EM signal plotted as a function of spread of the muon signal. We simulated 60 proton (blue) and 60 iron (red) showers with the QGSJet~II--04 model with energy $10^{19}$~eV and zenith angles 0$^{\circ}$ (left) and 45$^{\circ}$ (right). }
  \label{SemVsSmuQGSJetProton}
\end{figure}

Important aspects to be explored are the fluctuations of the muon and EM signals and their respective correlations. To include these effects in the simplified simulations of shower signals (Section~2.2), the correlations of $S_{\mu}^{19}$ and $S_{\rm EM}^{19}$ and their fluctuations were studied in detail for each zenith angle, for both models of hadronic interactions and for four primary particles. 
Relative fluctuations in the muon signal were estimated to be about 3\% for primary iron nuclei increasing up to 20\% for primary protons. It turned out that they change little in the whole range of zenith angles; the largest change was observed for protons (20\% at 0$^{\circ}$ and 15\% at 45$^{\circ}$). 
Somewhat smaller relative fluctuations in the EM signal occurred, in a way that they are reasonably well correlated with the relative muon fluctuations. 
However, the roughly linear relationship between the muon and EM signal depends on the zenith angle.
The properties of the signal fluctuations and their relationship are visualized in Fig.~\ref{SemVsSmuQGSJetProton}.
 
\subsection{Simplified Simulation of Shower Signals}
\label{ToyMC}
Utilizing characteristics of the reference signals described in Section~2.1, we performed simplified simulations of shower signals that are induced simultaneously in two idealized detector arrays responding to the muon and EM components, respectively. We produced a number of air showers for various mass composition scenarios preserving the basic properties of CORSIKA showers. These sets of air showers were characterized by their realistic energy spectrum. We assumed that the arrival directions of the primary particles are isotropic (uniform in $\cos^{2}(\Theta)$). Besides the zenith angle of its arrival, each shower was identified by the muon and EM signal that triggered simultaneous responses of the two arrays of detectors. The latter two quantities were obtained from the shower energy assuming the attenuation of secondary particles in the atmosphere and shower fluctuations supplemented by detector resolutions as
described in the following.

To be more specific, we used 286 four--component primary beams with protons and He, N and Fe nuclei. This way we cover all possible mass compositions which differ in the relative abundance of each of these four primaries in steps of 10\%. The shower energies were generated in the range ${\langle}10^{18.5},10^{20}{\rangle}$~eV using a spectral index 2.7 and including the GZK feature \cite{Greisen,ZatsepinKuzmin} at the end of the spectrum according to \cite{AUGER:ICRC13}. The shower zenith angles were generated randomly between 0$^{\circ}$ and 45$^{\circ}$ assuming they are distributed uniformly in $\cos^{2}(\Theta)$. The maximal value of the zenith angle corresponds to the upper bound for the full trigger efficiency of the EM detectors deployed in current arrays, see e.g.~\cite{TA:enespec}. 

In order to describe the detector response, we assumed that the shower energy is directly linked with a single signal that would be recorded when a shower of the same energy hits the detector array at a zenith angle of 38$^{\circ}$. 
Further we assumed that the shower energy $E_{\rm MC}$ and this shower signal, $S_{38}$, are related by
\begin{equation}
E_{\rm MC}=a\cdot S_{38}^{b} .
\label{eq1}
\end{equation}
Here we choose an arbitrary normalization constant $a=10^{16}$~eV. 
For the sake of simplicity the exponent $b=1$ was taken. 
It is not far from the value estimated at the Pierre Auger Observatory~\cite{AUGER:ICRC13}. 
It is worth noting that this relationship also reflects approximately the look--up table used at the Telescope Array experiment \cite{TA:enespec}.

In the subsequent analysis, the muon and EM signals were determined for a shower induced by a given primary particle with a given primary energy that is incident at a given zenith angle.
In the first step, we generated the type of the primary particle according to a chosen mass composition scenario, its energy and its zenith angle. The generated shower energy $E_{\rm MC}$ was transformed to the shower signal $S_{38}$ using Eq.~(\ref{eq1}). 
Then, muon and EM responses for the shower incident at the generated zenith angle were obtained.
For this purpose, we utilized corresponding reference shapes of the attenuation curves and their ratios (see Section~\ref{RefShowerSignals} and Figs.~\ref{CICfit},\ref{NucleiToProtonRatios} in particular) and applied them to the shower signal $S_{38}$.
In the next step, the fluctuations of the shower signals of both muon and EM components and their respective correlations were included into the analysis.
We used the results of our simulations described at the end of Section~\ref{RefShowerSignals} (see also Fig.~\ref{SemVsSmuQGSJetProton}).
Finally, another smearing of the muon and EM signals was additionally applied in order to model the effect of detector resolutions.
For both signals we used Gaussian smearing with a relative variances of 20\%.
Note that these detector resolutions were set to be worse than those resolutions quoted by the current detector arrays \cite{TA:enespec,AUGER:ICRC13}.
This way, we ended up with the simulated muon and EM signals, $S_{\rm EM}$ and $S_{\mu}$, that each air shower induces in the two idealized arrays of the muon and EM detectors at a wide range of primary energy preserving the basic properties of CORSIKA showers.
In particular, we simulated $7\times 10^{5}$ air showers for different mass compositions of the beam of the primary particles.

\subsection{Combined Signal Approach}
\label{sec:MatchedFraction}
In order to get sufficient information about the properties of the combined responses of the two arrays of detectors sensitive to the muon and EM secondaries, we studied different mass compositions of the primary beams. It turned out that the dispersion of the primary mass can be assessed by relying upon different shapes of the attenuation curves for the muon and EM signal, see also Fig.~\ref{NucleiToProtonRatios}. To quantify this finding, we introduced a parameter sensitive to the dispersion of the mass of a beam of primary particles causing air showers that generate the muon and EM signals in the two idealized arrays of detectors (see Section~2.2).

Following the procedure of inferring the CIC curve, our sets of MC data for each of the composition scenarios were each divided into 7 equidistant bins in $\cos^{2}(\Theta)$, where $0\leq \Theta \leq 45^{\circ}$, containing $\sim$100~000 events per bin. In each such bin, we selected $N_{\rm Cut}=12~000$ events with the highest signals induced in both the muon and EM detectors. This choice corresponds roughly to the selection of events with a primary energy higher than about $10^{19}$~eV. Specifically, for each bin of $\cos^{2}(\Theta)$, we selected a set $\rm M^{\mu}$ of 12~000 events with the highest signal in the muon detector and another set $\rm M^{\rm EM}$ of 12~000 events with the highest signal in the EM detector. Finally, we counted the number of identical events, $N_{\rm m} \rm = | M^{\mu} \cap M^{\rm EM} |$ $\leq N_{\rm Cut}$, that are present in both these event sets. 

\begin{figure}[h!]
  \centering
	\includegraphics[width=1.0\textwidth]{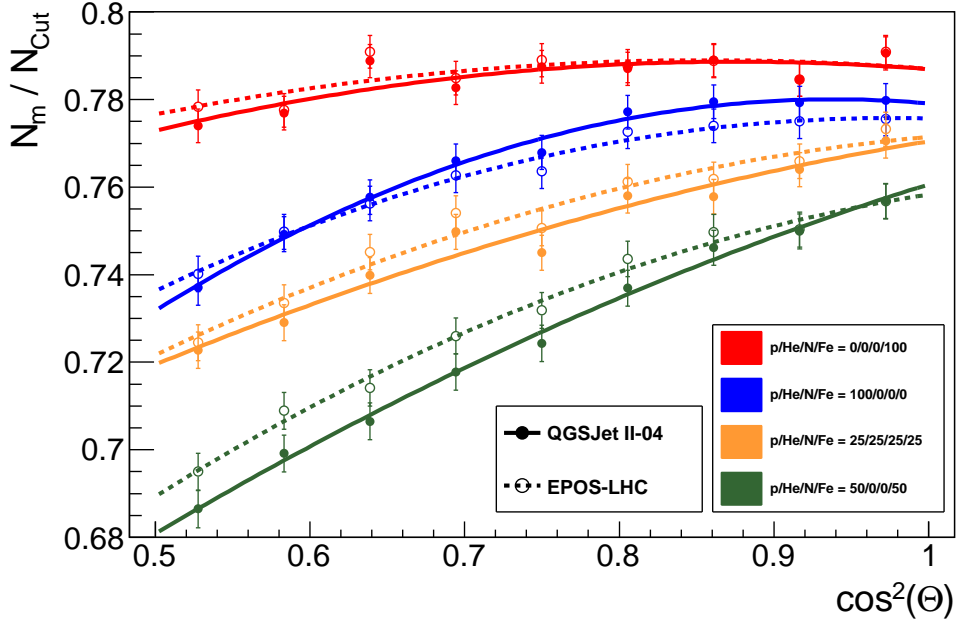}
  \caption{Attenuation of the matched fraction. Simulated fractions of events that are matched in both detectors, $N_{\rm m}/N_{\rm Cut}$, are plotted as a function of $\cos^{2}(\Theta)$. Results of four examples of the primary mass composition (different colors) and for the two models of hadronic interactions (full and empty markers) are shown. Depicted curves are quadratic fits to these fractions.}
  \label{MatchedNumber}
 \end{figure}

Dependencies of the fraction of the number of events that matched in both detectors, $N_{\rm m}/N_{\rm Cut}$, are shown in Fig.~\ref{MatchedNumber} as a function of zenith angle for four different primary beams. For any one--component scenario only a small mass dependent decrease of the number $N_{\rm m}$ is expected with increasing zenith angle. It is due to combined effects given by shower--to--shower fluctuations of the muon and EM signals, their correlations and the resolutions of both detectors (see results for pure proton beams in blue and for iron nuclei in red in Fig.~\ref{MatchedNumber}). On the other hand, as the primary particles become mixed, the orderings of events according to sizes of their signals in the two detectors diverge, resulting in an additional decrease of the number of matched events $N_{\rm m}$ with increasing zenith angle (see results for mixed compositions in orange and green in Fig.~\ref{MatchedNumber}). This effect is mainly caused by a very different ordering of the sizes of EM signal with respect to a primary particle type at different zenith angle bins. 

Let us consider, for example, that iron nuclei and proton primaries of similar energies incident at zenith angles of about 40$^{\circ}$ cause air showers that induce signals in the muon detector that are slightly larger than the signal $S_{\rm Cut}^{\mu}$. This signal corresponds to the chosen threshold number $N_{\rm Cut}$, i.e. these events still belong to the set $\rm M^{\mu}$. Consideration of these events was motivated by the fact that there is a higher chance for primary iron nuclei compared to primary protons to induce air showers that end up in the set of events $\rm M^{\mu}$, see left panel of Fig.~\ref{NucleiToProtonRatios} ($S_{\mu}^{19}$(Fe)$>S_{\mu}^{19}$(p) for $\Theta\sim40^{\circ}$). On the other hand, since the EM signal that is induced by primary iron nuclei in the EM detector is smaller on average than the EM signal caused by protons for considered events, see right panel of Fig.~\ref{NucleiToProtonRatios} ($S_{\rm EM}^{19}$(Fe)$<S_{\rm EM}^{19}$(p) for $\Theta\sim40^{\circ}$), there is a higher chance that the proton signal is larger than the threshold value $S_{\rm Cut}^{\rm EM}$ than it is for the iron signal. Here, the threshold signal $S_{\rm Cut}^{\rm EM}$ is determined by the threshold number $N_{\rm Cut}$ in the EM detector. Therefore most of the events under consideration that are caused by primary iron nuclei are not included in the set $\rm M^{EM}$. As a result, the number of matched events $N_{\rm m}$ decreases more steeply with the increasing zenith angle of incident particles for primary beams consisting of a mixture of particles than for pure primary beams, as demonstrated in Fig.~\ref{MatchedNumber}.

To a first approximation, the number of matched events $N_{\rm m}$ is well described by a quadratic function of $\cos^{2}(\Theta)$ for any primary composition. The fitted quadratic curves of the fraction of matched events are also shown in Fig.~\ref{MatchedNumber}. They were mostly found as decreasing functions of the increasing zenith angle. 
Inferred from these curves, we define a descriptive parameter that simply quantifies this dependence
\begin{equation}
\Phi = 1 - \frac{N_{\rm m} (\Theta = 45^{\circ})}{N_{\rm m} (\Theta = 0^{\circ})}.
\label{PhiEq}
\end{equation}
It expresses the decrease of the number of events $N_{\rm m}$ matched in both muon and EM detectors with zenith angle from $0^{\circ}$ to $45^{\circ}$ as displayed in Fig.~\ref{MatchedNumber}. For the pure primary beams, the parameter $\Phi$ acquires larger values for lighter primaries. Even larger values are expected for the non--zero variance of the mass of incident primaries.

\section{Sensitivity to Mass Composition}
\label{sec:MassFluct} 
In what follows, we examined several sets of four--component primary
beams. These beams were characterized by the mean and variance of the logarithmic mass of primary nuclei. Specifically, we assumed that primary cosmic rays consist of four nuclei with $A_{i}$ nucleons, $i$ = 1, 2, 3, 4, contributing with relative abundances $f_{i}~\in$~${\langle}0,~1{\rangle}$, where $\sum\limits^4_{i=1} f_{i}=1$. The dispersion of the logarithmic mass in the primary beam, $\sigma^2(\ln A)$, is given by
\begin{equation}
\sigma^2(\ln A) = \sum\limits^4_{i=1} f_{i}\cdot(\ln A_i -\langle \ln A \rangle)^2, \;\;\;\;\;\;\;\;\; \langle \ln A \rangle = \sum\limits^4_{i=1} f_{i}\cdot\ln A_{i}.
\label{FluctEq}
\end{equation}
where $\langle \ln A \rangle$ is the mean of the primary logarithmic mass.

\subsection{Dispersion of Primary Masses}
Using the two models of hadronic interactions, we simulated the muon
and EM signals as they occur in our idealized arrays of detectors for each of the 286 arbitrarily chosen four--component primary beams within the simplified treatment described in Section~\ref{ToyMC}. We determined corresponding matched fractions of the muon and EM events as functions of the zenith angle, see Section~\ref{sec:MatchedFraction}. Finally, for each of the chosen primary composition we determined the descriptive parameter $\Phi$ given in Eq.~(\ref{PhiEq}), and we related this parameter with the given dispersion of the logarithmic mass in the primary beam.

\begin{figure}[h!p]
  \centering
\includegraphics[width=0.85\textwidth]{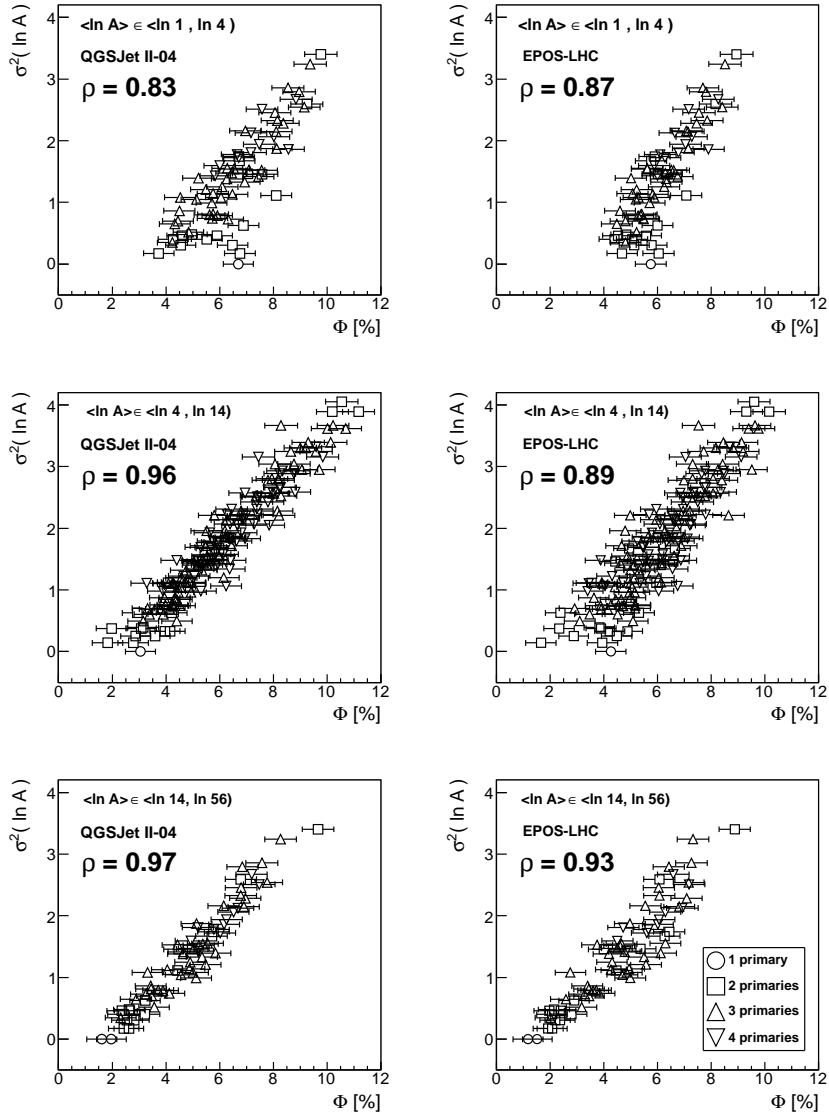}
  \caption{Sensitivity to the dispersion of primary masses. The variance of the logarithmic mass of the primary beam, $\sigma^2(\ln A)$ in Eq.~(\ref{FluctEq}), is plotted as a function of the parameter $\Phi$ defined in Eq.~(\ref{PhiEq}) for QGSJet~II--04 (left panels) and EPOS--LHC (right panels). Three regions of the mean logarithmic mass ($\langle \ln A \rangle$) are selected (rows of panels) from 286~different mass compositions of the primary beam. The Pearson's correlation coefficient $\rho$ is calculated for each of 6 plots. Markers indicate the number of primaries present in the beam.}
  \label{FluctMassRelation}
 \end{figure}

Our results for three regions of $\langle \ln A \rangle$ are summarized in Fig.~\ref{FluctMassRelation}. In this figure, the dispersion of the primary beam is depicted as a function of the parameter $\Phi$ for QGSJet~II--04 (left panels) and EPOS--LHC (right panels). We analyzed the incident beams with a single primary up to four primary components. Good agreement is observed between the results obtained with both examined models of hadronic interactions. In our analysis, the spread of the primary beam masses increases with the difference of the number of matched events $N_{\rm m}$ between the zenith angles of $0^{\circ}$ and $45^{\circ}$ (see Fig.~\ref{MatchedNumber}), as described by the parameter $\Phi$ introduced in Eq.~(\ref{PhiEq}). The Pearson's correlation coefficient ($\rho$) of $\Phi$ and $\sigma^2(\ln A)$ is calculated for each region of $\langle \ln A \rangle$ and each model of hadronic interactions.

It turns out that the parameter $\Phi$ explains well the dispersion of logarithmic mass of the primary beam constituents. Obviously, the knowledge of the mean logarithmic mass increases its explanatory power. To demonstrate this feature, we use different bins of $\langle \ln A \rangle$ of a width of $w\approx 1.3$. It is equivalent to the information that $\langle \ln A \rangle$ can be obtained from an independent measurement at the same energy. The choice of $w$ is realistic since, for example, the uncertainty derived from the measurements performed at $10^{19}$~eV is typically $\delta_{\rm ln A}=0.4< w/2$ \cite{AUGER:LongXmaxPaper}. 

On average, the parameter $\Phi$ decreases with the increasing mean logarithmic mass. Also, the correlation between $\Phi$ and $\sigma^2(\ln A)$ is improving with the increasing mean logarithmic mass.

It is worth emphasizing that the parameter $\Phi$ behaves similarly with the dispersion of the primary mass for a wide range of the selected thresholds $N_{\rm Cut}$. In our examples, $N_{\rm Cut}=12~000$ is only a matter of an arbitrary choice reflecting the total number of events we have simulated and the width of the $\cos^{2}(\Theta)$ bin we have chosen. Also the numbers of matched events, $N_{\rm m}$, are not of crucial importance in our treatment, only their relative changes with the zenith angle play any roles.

In our procedure, the size of the number of matched events is primarily given by the shower--to--shower fluctuations and by the resolutions of both detectors. On the other hand, the evolution of the matched number with the zenith angle is mainly caused by corresponding reference responses of the muon and EM detectors as obtained for different primary nuclei in Section~\ref{RefShowerSignals}.

\subsection{Discussion}
The presented relationship between the variance of the logarithmic mass of the primary beam, $\sigma^{2} (\ln A)$, and the parameter $\Phi$ derived from different responses of the two arrays of different detectors is rather general. It stems from the basic properties of available shower observables as well as from the adopted CIC approach. Somewhat different detector responses that might be various functions of the zenith angle, while proportional to the muon or EM density, would not change our results significantly. 

For example, the muons were not considered in the response of the EM detector (thin scintillator) since their inclusion cannot change our results substantially. The reason is that the ground density of the total number of muons is about 15--50 times smaller than the ground density of EM particles. Moreover, the zenith angle behaviour of the muon component is rather moderate.

Nonetheless, we stress that a specific application of the parameter $\Phi$ to quantify the spread of mass in the primary cosmic ray beam for the measured UHECR data would require a detailed knowledge of the detectors' responses. Taking into account the specificity of their observables, this technique could also be generalized for observatories studying cosmic rays of lower energies.

In a sense, the effects of shower--to--shower fluctuations are not substantial in our treatment. We verified numerically that the results presented in Fig.~\ref{FluctMassRelation} will change only marginally for several times stronger correlations between the muon and EM signals. Less significant fluctuations and correlations incorporated in simulated signals leave our results almost unaffected. The studied relationship between $\sigma^{2}(\ln A)$ and $\Phi$ remains valid also in cases where the shower signals are taken at various distances from the shower core. The same holds for different threshold energies of detected secondary particles in both types of detectors.

The crucial ingredient of this work is that the two most up--to--date models of hadronic interactions do not show any substantial deviations in terms of shapes of reference attenuation curves and their ratios. For a given primary composition, the relationship between the two types of signals, as expressed by the parameter $\Phi$, does not depend strongly on the actual values of the muon and EM signals. Therefore, the details of the models of hadronic interactions are rather suppressed in our treatment. One needs to keep in mind, however, that the properties of the reference showers described in Section~\ref{RefShowerSignals} could still be different from the properties of the real showers detected in the current detector arrays.

In comparison with the statistical uncertainty of the currently most precise method based on the depth of shower maximum measured at $10^{19}$~eV \cite{AUGER:LongXmaxPaper}, the spread of the mass $\sigma^{2} (\ln A)$ can be determined with a similar uncertainty $\delta_{\sigma^2(\rm ln A)}\simeq0.5$. This uncertainty was estimated as the variance of the distribution of $r=\sigma^{2}_{\rm meas} (\ln A)-\sigma^{2} (\ln A)$, where $\sigma^{2}_{\rm meas}$ was calculated using a linear fit of $\sigma^{2} (\ln A)$ depending on $\Phi$ for different regions of $\langle \ln A \rangle$ and various models of hadronic interactions. The variance of $r$ is decreasing with increasing $\langle \ln A \rangle$, see also correlation coefficients in Fig.~\ref{FluctMassRelation}. Similar or even better precision, when compared to other methods, and weak dependence on the model assumption make the presented method advantageous to complement frequently conducted studies of the mass composition that are based on the analysis of the mean logarithmic mass of primary species.

\section{Conclusions}
The basic properties of CORSIKA showers incorporated in the two post--LHC models of hadronic interactions were used for simplified simulations of the response of a combined detector of UHECR. In particular, the attenuation curves and fluctuations of electromagnetic and muon densities registered at ground level were studied together with their correlations. We applied the principle of the constant intensity cut method to two different types of arrays sensitive to the electromagnetic and muonic signals, respectively, and combined their outputs. We demonstrated that the dispersion of the mass in the primary beam of the UHECR particles can be measured using the zenith angle behaviour of the number of matched events in both types of arrays. We obtained very similar results for the two selected models of hadronic interactions.

\section*{Acknowledgments}
We would like to acknowledge valuable interactions with our colleagues of the Pierre Auger Collaboration. We are very grateful to St\'{e}phane Coutu for careful reading of our manuscript and his suggestions. We also appreciate the help of Vladim\'ir Novotn\'y and Stanislav \v{S}tef\'anik. This work is funded by the Czech Science Foundation grant 14-17501S.







\end{document}